\documentclass[aps,prb,twocolumn,superscriptaddress,showpacs]{revtex4-1}

\usepackage{hyperref}
\usepackage[dvips]{graphicx}
\usepackage{amssymb,amsmath,amsfonts}

\begin{document}


\title{Reversible modifications of linear dispersion - graphene between boron nitride monolayers}


\author{J. S\l awi\'{n}ska}
\affiliation{Theoretical Physics Department II, University of Lodz, Pomorska 149/153, 90-236 Lodz, Poland}

\author{I. Zasada}
\affiliation{Solid State Physics Department,University of Lodz, Pomorska 149/153, 90-236 Lodz, Poland}

\author{P. Kosi\'{n}ski}
\affiliation{Theoretical Physics Department II, University of Lodz, Pomorska 149/153, 90-236 Lodz, Poland}

\author{Z. Klusek}
\affiliation{Solid State Physics Department,University of Lodz, Pomorska 149/153, 90-236 Lodz, Poland}


\vspace{.15in}

\begin{abstract}
Electronic properties of the graphene layer sandwiched between two hexagonal boron nitride sheets have been studied using the first principles calculations and the minimal tight-binding model. It is shown that for the ABC-stacked structure in the absence of external field the bands are linear in the vicinity of the Dirac points as in the case of single layer graphene. For certain atomic configuration the electric field effect allows opening of a bandgap of over 230 meV. We believe that this mechanism of energy gap tuning could significantly improve the characteristics of graphene-based field effect transistors and pave the way for novel electronic applications.
\end{abstract}

\maketitle


\section{Introduction}
Graphene, the thinnest electronic material with amazing properties\cite{novoselov_science} is a promising candidate for next-generation electronics. Its cone-shaped valence and conduction bands touch each other at the corners of the Brillouin zone. The large area graphene is then semimetallic and not suitable for use in digital logic devices due to the fact that such channels cannot be switched off. However, the band structure of graphene can be modified and many approaches have been proposed to break the symmetry between sublattices responsible for massless character of Dirac fermions and the lack of the bandgap.\cite{review} Theoretical investigations have demonstrated that a narrow bandgap can be created by placing graphene sheet on the lattice-matched materials, such as hexagonal boron nitride (h-BN).\cite{hbn} An alternative route to induce the bandgap is functionalization of graphene with suitable elements, as fluorine (fluorination\cite{fluor}) and hydrogen (hydrogenetion\cite{hydro}), which transforms graphene into an insulator with bandgap of about 3.5 eV. This process is reversible, but it takes hours to regenerate pristine graphene from these materials.

On the other hand, energy gap tuning by electric field effect in bilayer Bernal stacked graphene has been widely studied.\cite{mccann, castro_prl} Recently, a bilayer graphene transistor with a sizable on/off current ratio has been fabricated\cite{nanoletters} showing a great potential of graphene-based electronics. One of the main difficulties of producing bilayer graphene transistors is disorder. For example, graphene devices on SiO$_{2}$ substrate are highly disordered due to the scattering from charged surface states and impurities which limits the carrier mobility and suppresses the graphene unique properties. The measured moblities 10000-15000 cm$^{2}$V$^{-1}$s$^{-1}$ are much smaller than those reported for suspended graphene.
However, the very recent experimental studies have revealed\cite{substrat_hbn} that graphene transferred onto h-BN substrate is free of wrinkles, distortions and surface charge traps. This leads to the construction of graphene devices with very high mobility close to 60000 cm$^{2}$V$^{-1}$s$^{-1}$.

The conventional semiconductors exhibit a general trend that with increasing bandgap, the electron mobility decreases.\cite{schwierz} This behavior has been also predicted for carbon nanotubes and graphene nanoribbons. In both above mentioned cases, the valence and conduction bands become progressively parabolic (rather than linear) when the larger bandgap opens. It causes the increase in the effective masses of charge carriers and, in consequence, the decrease in their mobility. In the unbiased bilayer graphene the low energy bands are parabolic, not cone-shaped, which lead to a finite value of the effective mass. If the linear dispersion of pristine graphene was preserved in the tunable system, it would offer new possibilities for electronics based on the properties of massless Dirac fermions. In particular, the high carrier mobility in graphene at room temperature would profit without the loss of the possibility to switch devices off.

\begin{figure*}
\includegraphics{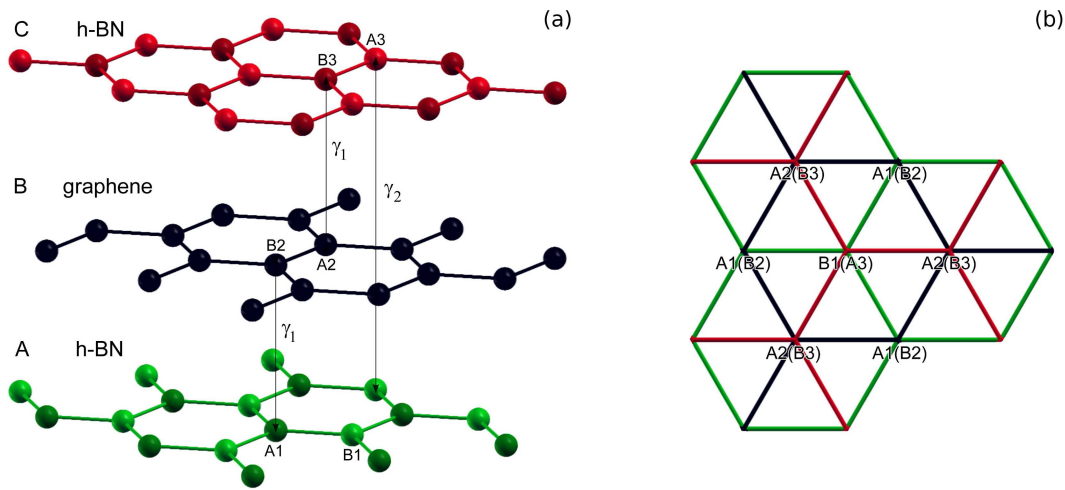}
\caption{\label{uklad}(Color online) (a) Lattice structure of h-BN/graphene/h-BN ABC-stacked trilayer system, darker and lighter balls denote atoms of the same type in h-BN layers (b) orientations of graphene and h-BN layers with respect to each other.}
\end{figure*}

Our recent theoretical studies\cite{nasze} on graphene/h-BN systems have revealed that the energy gap tuning in the range starting from 0 to 130 meV can be achieved in graphene/h-BN bilayer system. However, the linear dispersion is not exactly preserved - the mass of Dirac fermions is small, but cannot be reduced to zero.

In this paper we report on the possibility of creating and tuning the bandgap in trilayer systems consisting of graphene monolayer sandwiched between two h-BN honeycombs (see Fig. \ref{uklad}). It seems that in ABC stacked trilayer h-BN/graphene/h-BN system the presence of two identical boron-carbon or nitrogen-carbon dimers in the unit cell does not induce the high asymmetry between sublattices, which guarantees the conical dispersion exactly preserved in the K points. When the electric field is applied perpendicularly to the layers, the bands are shifted and deformed which leads to the bandgap opening.

\section{Geometry of trilayers}

We have studied a stability of graphene/h-BN AB-stacked bilayers and h-BN/graphene/h-BN trilayers with ABA and ABC stacking orders by means of density functional theory (DFT) calculations\footnote{DFT calculations were performed in the local density approximation (LDA), using Perdew-Zunger exchange-correlation functional\cite{perdewzunger} and the norm conserving pseudopotentials. We used PWscf package\cite{pwscf} of the Quantum Espresso distribution.\cite{qe} The plane wave kinetic energy cutoff was set at 400 Ry. The Brillouin zone was sampled using a 36 $\times$ 36 $\times$ 1 grid with the tetrahedron method.\cite{tetrahedron} 
We found equilibrium distances between the layers by performing a set of total energy calculations for different values of separations assuming that positions of atoms in each layer are fixed. We used one in-plane lattice constant for both materials equal to $a$=2.456 \AA. The interactions with periodic images of the supercell in the z-direction were avoided by setting a large vacuum region of 13 \AA. The electric field was simulated by a sawlike potential. Total energies were convergent within to $10^{-4}$ Ry.}. In case of the bilayer there are two inequivalent orientations of the graphene with respect to the h-BN lattice. In the lowest-energy configuration carbon atoms $B_{2}$ and $A_{2}$ are placed directly above the boron atom ($A_{1}$) and on top of the empty center of the hexagon in the bottom layer, respectively. In the second configuration, the boron and nitrogen atoms are exchanged, thus the unit cell contains a nitrogen-carbon (N-C) dimer. 

The relative position of h-BN and graphene layers allows for two different orientations of the third layer in the sandwich. The same orientation of both h-BN layers gives ABA Bernal stacking configuration typical for graphite. In the ABC (rhombohedral) stacking (see Fig. \ref{uklad} (b)) each pair of nearest-neighboring layers forms an AB-stacked bilayer with the upper B sublattice directly on top of the lower A sublattice and the upper A above the center of the hexagon below.\cite{macdonald} The optimized distances between the layers are the same for ABA and ABC stacking orders. The difference in total energies is less than 0.1 meV which is within a computational error. Moreover, it seems that the particular configuration would be determined by the substrate, as in the case of the graphene trilayers on SiC.\cite{sic} In case of two inequivalent configurations containing N-C or B-C dimers, there is a larger difference in total energies and interlayer separations, but both of them are stable. The energy difference is then over 30 meV preferring the latter configuration, while the distances between the layers are equal to 3.44 and 3.22 \AA, respectively.

\section{Tight binding approach}
We have analyzed the electronic properties of h-BN/graphene/h-BN trilayers using the nearest-neighbor tight-binding model. We include only the interlayer hopping between atoms which are on top of each other in adjacent layers, thus the interaction between the sheets takes place through half of the atoms in each plane. This method has been previously applied to study trilayer graphene and allowed for simple and qualitatively correct analysis of this system, with and without the applied external fields.\cite{uz1}

First, we remind the hamiltonian describing the bottom bilayer in the system. We include the parameter $U$ of the interlayer asymmetry induced by an external field, which is uniform and perpendicular to the layers. In case of the AB-stacked bilayer we construct the hamiltonian in the basis with atomic components $\Phi_{A_{1}}$, $\Phi_{B_{1}}$, $\Phi_{A_{2}}$, $\Phi_{B_{2}}$ (only $2p_{z}$ orbitals of respective atoms are taken into account, that is $\Phi_{A_{1}}$ is B $2p_{z}$ orbital with Bloch factor, $\Phi_{B_{1}}$ N $2p_{z}$, $\Phi_{A_{2}}$, $\Phi_{B_{2}}$ are $2p_{z}$ orbitals of carbon atoms with Bloch factors) \cite{nasze}:

\begin{equation}H_{AB}=\left( \begin{array}{cccc}\vspace{0.1in}
\epsilon^{\alpha}+\frac{U}{2}& \gamma_{0}'f(\vec{k})&0&\gamma_{1}\\\vspace{0.1in}
\gamma_{0}'f(\vec{k})^{*} & \epsilon^{\beta}+\frac{U}{2}&0&0\\\vspace{0.1in}
0& 0&-\frac{U}{2}&\gamma_{0}f(\vec{k})\\
\gamma_{1}& 0&\gamma_{0}f(\vec{k})^{*}&-\frac{U}{2}\end{array} \right)\end{equation}

where we fixed the zero energy as $2p$ level of carbon atoms in the crystal. 

As in the Slonczewski-Weiss-McClure model of the AB bulk graphite, $\gamma_{0}'$ ($\gamma_{0}$) and $\gamma_{1}$ describe the nearest-neighbor intralayer and interlayer hopping parameters respectively, $\epsilon^{\alpha}$ and $\epsilon^{\beta}$ denote onsite energies of inequivalent atoms in the h-BN honeycomb and $f(\vec{k})=\sum_{i=1}^{3}e^{i\vec{k}\vec{\delta}_{i}}$, where $\vec{\delta}$ stands for the in-plane nearest-neighbor vectors.

We generalize this minimal model on the case of the ABC trilayer including $\Phi_{A_{3}}$ and $\Phi_{B_{3}}$ components and consider the case when the lower layer is at potential U/2, the upper layer is at potential -U/2 and the middle is at zero potential:
\small
\begin{equation}\begin{pmatrix}
\epsilon^{\alpha}+\frac{U}{2}& \gamma_{0}'f(\vec{k})&0&\gamma_{1}&0&0\\
\gamma_{0}'f(\vec{k})^{*} & \epsilon^{\beta}+\frac{U}{2}&0&0&0&0\\
0& 0&0&\gamma_{0}f(\vec{k})&0&\gamma_{1}\\
\gamma_{1}& 0&\gamma_{0}f(\vec{k})^{*}&0&0&0\\
0 & 0&0&0&\epsilon^{\beta}-\frac{U}{2}&\gamma_{0}'f(\vec{k})\\
0& 0&\gamma_{1}&0&\gamma_{0}'f(\vec{k})^{*}&\epsilon^{\alpha}-\frac{U}{2}\\
\end{pmatrix}\normalsize\end{equation}
\normalsize
where we neglected all hopping parameters between non-adjacent layers.\footnote{DFT results show that the hopping parameter $\gamma_{2}$ between $B_{1}$ and $A_{3}$ atoms lifts the degeneracy of $E_{3}=\epsilon^{\beta}\pm\gamma_{2}$. Moreover, the minimal tight-binding model does not well reproduce the crossed pair of bands $E_{1}$ or $E_{2}$ depending on B-C or N-C configuration, respectively. The highest valence and the lowest conduction bands are precisely described by a minimal model with the determined set of parameters. In all cases the bands are well described qualitatively.}.

It is worth to note that the above hamiltonian resembles that of ABC graphite\cite{uz3} and graphene trilayer\cite{uz2}. However, the system contains only one graphene layer and its properties are more similar to graphene monolayers rather than to trilayers.

In the corners of the Brillouin zone, $|f|=0$, thus the eigenvalues can be easily determined

\begin{equation}E_{1}^{\pm}=\frac{\epsilon^{\alpha}}{2}\pm\frac{1}{4}U+\sqrt{\gamma_{1}^{2}+\biggl(\frac{U\pm2\,\epsilon^{\alpha}}{4}\biggr)^{2}}\end{equation}
\begin{equation}E_{2}^{\pm}=\frac{\epsilon^{\alpha}}{2}\pm\frac{1}{4}U-\sqrt{\gamma_{1}^{2}+\biggl(\frac{U\pm2\,\epsilon^{\alpha}}{4}\biggr)^{2}}\end{equation}
\begin{equation}E_{3}^{\pm}=\epsilon^{\beta}\pm\frac{U}{2}\end{equation}

\begin{figure}
\includegraphics{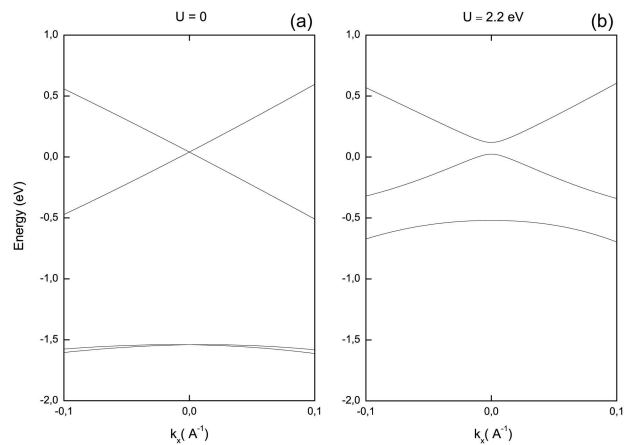}
\caption{\label{wykresy} TB band structure of h-BN/graphene/h-BN ABC-stacked trilayer with N-C dimers in the absence (a) and in the presence (b) of the external electric field perpendicular to the layers.
}
\end{figure}

One can observe that in the absence of the external field ($U=0$) there are three double roots of the secular equation. It means that there is no energy gap between valence and conduction bands, which is clearly visible in Fig. \ref{wykresy} (a). DFT calculations have shown that the degeneracy is in part broken by next-nearest neighbor interlayer hopping between sites $A_{1}$ and $B_{3}$ that lie on the same vertical line\cite{uz4}, but it does not affect the highest valence and the lowest conduction bands.

Moreover, it is easy to show by the methods of the perturbation theory that there are two pairs of bands, which are linear in the vicinity of the Dirac points:

\begin{equation}E_{1,2}(\vec{k})=E_{1,2}\pm\frac{\gamma_{0}|f(\vec{k})|(\frac{\gamma_{1}}{E_{1,2}})^{2}}{1+(\frac{\gamma_{1}}{E_{1,2}})^{2}}\end{equation}
since $|f(\vec{k})|$ is linear there. It has to be stressed that the system is half-filled and the Fermi level coincides with the touching point of the cones.

\begin{figure}
\includegraphics{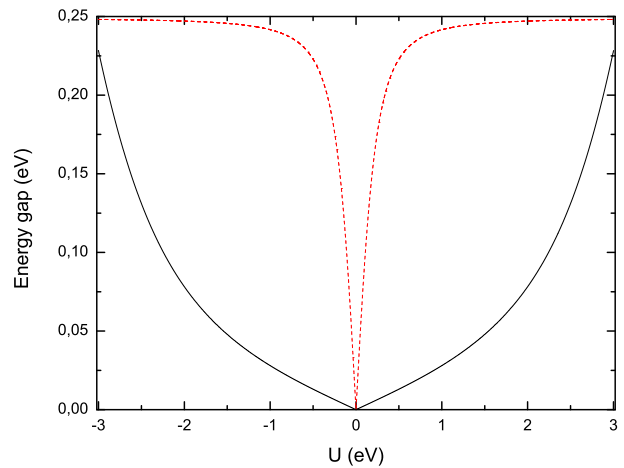}
\caption{\label{przerwa} (Color online) Variation of the energy gap as a function of U in the range, where the gap is direct. The case of ABC-stacked trilayer containing N-C dimers is chosen. Red (dashed) line denotes the dependence for biased graphene bilayer. For simplicity the same value of interlayer hopping $\gamma_{1}$ is taken for both systems.
}
\end{figure}

The valence and conduction bands are linear in the wide range of energies ($\pm$1 eV with respect to the touching point) with the value of Fermi velocity nearly equal to that of graphene. The external electric field ($U\neq0$) lifts the degeneracy in the K point leading to shifts and the repulsion of bands (see Fig. \ref{wykresy} (b)). The variation of the induced energy gap between valence and conduction bands is shown in Fig. \ref{przerwa} and compared to the bandgap opening in biased bilayer graphene.

\section{Numerical results and discussion}

\begin{figure}
\includegraphics{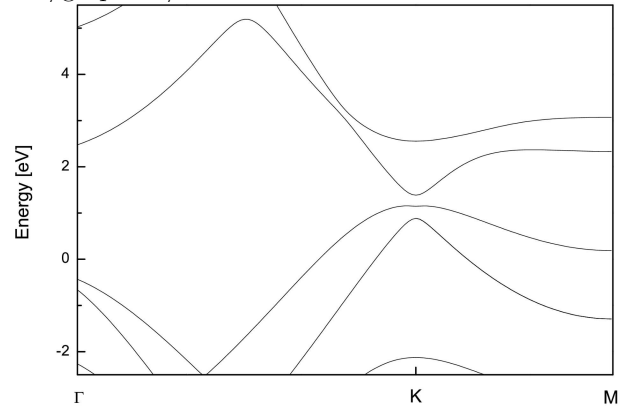}
\caption{\label{pasma}DFT band structure along $\Gamma$KM line in momentum space of trilayer h-BN/graphene/h-BN in the presence of external electric field applied in the direction perpendicular to the layers. The indirect energy gap is 231.5 meV.
}
\end{figure}

The ground state configuration containing B-C dimer does not offer the possibility of energy gap tuning in the sufficiently large range, which is related to the onsite energies values. Among h-BN/graphene/h-BN sandwiched trilayers, the ABC-stacked system containing nitrogen-carbon dimers seems to have the largest potential for applications. We have found the approximate values of the tight-binding parameters describing this system.\footnote{The TB band structure was fitted to DFT data in the straight vicinity of the K point along $\Gamma K$ line using the least squares method.} The onsite energies values are $\epsilon^{\alpha}=-1.50$ eV and $\epsilon^{\beta}=3.16$ eV representing nitrogen and boron $2p$ levels in the crystal lattice. Intralayer hopping parameters are equal to $\gamma_{0}'$=2.28 eV and $\gamma_{0}$=2.59 eV for h-BN and graphene layers, respectively. The value of an out-of-plane hopping has been estimated to $\gamma_{1}$=0.25 eV. According to the predictions of TB model with this set of parameters, the largest energy gap that can be created between the valence and conduction bands is nearly 230 meV and strongly depends on the $\gamma_{1}$ value. DFT results show that the bandgap achieves 232.6 meV suggesting that the parameters slightly change in biased system. Figure \ref{pasma} illustrates DFT predictions for the band structure of the system with the effective value of U exceeding 3 eV, where a large bandgap is opened. 

\section{Perspectives}
In summary, we have explored that in the ABC- stacked h-BN/graphene/h-BN trilayer one can observe the conical energy dispersion relation near the Dirac point, which is typical for pristine graphene. The electric field effect allows to create an energy gap up to about 230 meV depending on the exact value of interlayer hopping parameter $\gamma_{1}$. 

We proposed a unique mechanism of bandgap tuning in graphene-based system, which, under certain conditions, could be an alternative for graphene bilayer on SiO$_{2}$. Figure 3 illustrates that in the considered trilayer the gap opens slowly as compared to bilayer graphene. DFT results show that the screening effects are of the same order as in the previously studied graphene/h-BN bilayer\cite{nasze} as well as in the bilayer graphene\cite{hongkimin}, thus, in practice, the opening of sizable bandgap can be difficult. On the other hand, the disorder which is partly responsible for difficulties of producing bilayers transistors, in graphene/h-BN systems is significantly reduced indicating the very high quality of this substrate.\cite{substrat_hbn} In the present work, we have studied the issue of the band structure modifications. It is very important, but not the only factor that determines the characteristics of graphene FETs. We believe that the advantage of linear dispersion preserved in the unbiased trilayer can improve current on/off ratios or the speed of graphene devices\cite{schwierz}, however, a further theoretical analysis of the real systems and experimental verification have to be performed. Preliminary results on fabrication of graphene/h-BN combined structures have been recently reported,\cite{substrat_hbn, nowy_nature} thus we hope that our work will encourage experimentalists to produce h-BN/graphene/h-BN stacks.

\begin{acknowledgements}
This work is financially supported by Polish Ministry of Science and Higher Education in the frame of Grant No. N~N202~204737. One of us (J.S.) acknowledges support from the European Social Fund and Budget of State implemented under the Integrated Regional Operational Program, Project: Scholarship supporting postgraduate students' innovative scientific research and from the European Social Fund implemented under the Human Capital Operational Programme (POKL), Project: D-RIM. XCrysDen\cite{kokalj} was used to prepare Fig. \ref{uklad}.
\end{acknowledgements}
\providecommand{\noopsort}[1]{}\providecommand{\singleletter}[1]{#1}%
\end{document}